# THERMAL INTERACTION BETWEEN FREE CONVECTION AND FORCED CONVECTION ALONG A VERTICAL CONDUCTING WALL

Jian-Jun SHU[*] and I. Pop
School of Mechanical & Aerospace Engineering,
Nanyang Technological University,
50 Nanyang Avenue,
Singapore 639798,
E-mail: mjjshu@ntu.edu.sg

---

[*] Corresponding author.
E-mail address: mjjshu@ntu.edu.sg (J.-J. Shu).

## ABSTRACT
A theoretical study is presented in this paper to investigate the conjugate heat transfer across a vertical finite wall separating two forced and free convection flows at different temperatures. The heat conduction in the wall is in the transversal direction and countercurrent boundary layers are formed on the both sides of the wall. The governing equations of this problem and their corresponding boundary conditions are all cast into a dimensionless form by using a non-similarity transformation. These resultant equations with multiple singular points are solved numerically using a very efficient singular perturbation method. The effects of the resistance parameters and Prandtl numbers on heat transfer characteristics are investigated.

## INTRODUCTION

The study of thermal interaction between two semi-infinite fluid reservoirs at different temperatures through a vertical conductive wall is a very important topic in heat transfer because of its numerous engineering applications. This heat transfer process applies to reactor cooling, heat exchangers, thermal insulation, nuclear reactor safety, *etc*. Additionally, such interaction mechanism is, for the most part, inherent in the design of heat transfer apparatus. On the other hand, it is worth mentioning that demands in heat transfer engineering have requested researchers to develop the new types of equipments with superior performances, especially compact and lightweight ones. The need for small-size units requires detailed studies on the effects of interaction between the thermal field in both fluids and the wall conduction, which usually degrades the heat exchanger performance.

## NOMENCLATURE

| | | |
|---|---|---|
| $A,C$ | [-] | Constant |
| $b$ | [m] | Thickness of plate |
| $f$ | [-] | Reduced stream function |
| $g$ | [m/s$^2$] | Acceleration due to gravity |
| $h$ | [W/m$^2$K] | Heat transfer coefficient |
| $k$ | [W/mK] | Thermal conductivity |
| $L$ | [m] | Length of plate |
| $N_u$ | [-] | Nusselt number |
| $P_r$ | [-] | Prandtl number |
| $q,Q$ | [W/m$^2$] | Heat flux |
| $R_e$ | [-] | Reynolds number |
| $R_t$ | [-] | Forced convection thermal resistance parameter |
| $R_t^*$ | [-] | Free to forced convection parameter |
| $t,T$ | [K] | Temperature |
| $\Delta t$ | [K] | Characteristic ambient temperature |
| $u,v$ | [m/s] | Velocity components |
| $U_\infty$ | [m/s] | Free stream velocity |
| $x,y$ | [m] | Cartesian coordinates |

Special characters

| | | |
|---|---|---|
| $\alpha$ | [m$^2$/s] | Thermal diffusivity |
| $\beta$ | [1/K] | Coefficient of thermal expansion |
| $\theta$ | [-] | Dimensionless temperature |
| $\lambda$ | [-] | Dummy variable |
| $\nu$ | [m$^2$/s] | Kinematic viscosity |
| $\xi, \eta$ | [-] | Reduced coordinates |
| $\psi$ | [m$^2$/s] | Stream function |

Subscripts

| | |
|---|---|
| $c$ | Cold fluid system |
| $h$ | Hot fluid system |
| $s$ | Solid wall |
| $w$ | Condition at wall |
| $x$ | Local variable |

The problem of heat exchange between two free convection systems separated by a finite vertical conductive wall has not been studied extensively because of difficulties in solving the developments of flow and thermal boundary layers simultaneously. The local similarity method was applied to investigate theoretically the effect of thermal coupling produced by conduction through a finite vertical wall separating two free convection systems [1]. The study was done under the assumption that the wall conduction is only in the transversal direction. The governing boundary layer equations were transformed by introducing semi-similar variables and then solved numerically using a finite-difference method. This problem was also treated by using a more simple analysis based on a superposition method and conducting interferometric experiments to confirm the validity of the approximated theoretical results [2]. The modified Oseen linearized method [3,4] was employed to solve such a conjugate problem for the very large Prandtl number [5]. It was shown that the overall heat transfer rate was relatively independent by the Prandtl number. The problem of coupled heat transfer between two free convection systems separated by a finite vertical conducting wall was studied based on assuming the two-dimensional conduction equation in the wall [6]. In other words, the wall conduction takes place in both axial and transversal directions. Numerical solutions for both free convection systems and the analytical solution for the wall conduction were combined to obtain final solutions for the flow and heat transfer characteristics which fit the conjugate boundary conditions at both sides of the wall. Experiments were also conducted for the air-air systems with the conducting wall made of aluminium or glass. It was found that theoretical results described well the experimental temperature distributions. The numerical and asymptotic solutions of the free convection boundary layers were reported on both sides of a vertical conducting wall for all possible values of two main parameters [7].

It is noticed to this end that there were also published several papers on the problem of thermal interaction between the laminar film condensation of a saturated vapor and a forced or free convection system separated by a vertical conducting wall [8-12].

In this paper a new theoretical method is proposed to predict the heat transfer between free convection on one side of a finite vertical conducting wall and forced convection flow on the other side of the wall with the consideration of the wall thermal resistance. The conduction in the wall is in the transversal direction. Since both the plate temperature and the heat flux through the plate are unknown a priori in this problem, the boundary layer equations on both sides of the wall and the one-dimensional heat conduction equation for the wall are solved simultaneously. The numerical method used is a new comprehensive and non-iterative scheme based on the singular perturbation method [13,14]. This method differs by the iterative guessing technique proposed in [15,16]. Heat transfer characteristics have been derived for some main parameters entering this problem.

## BASIC EQUATIONS

The physical model under consideration along with the coordinate systems is shown in Figure 1, where the vertical plate with length $L$ and thickness $b$ separates two semi-infinite fluid reservoirs at different temperatures. The warmer reservoir contains a stagnant fluid with temperature $t_h$, while the ambient temperature on the cold side of the plate is $t_c$. Obviously, $t_h$ is higher than $t_c$. The upper left corner of the plate coincides with the origin of a Cartesian coordinate system whose $y$ axis points in the direction normal to the plate, while the $x$ axis points downward in the plate's longitudinal direction. Due to gravity, a free convection laminar boundary layer appears on the hot side of the plate and flows downward along the plate. A forced convection flow of the cooling fluid with velocity $U_\infty$ is imposed on the right lateral surface of the plate thus generating a forced convection boundary layer on this surface, which develops with increasing thickness downstream. Accordingly, two fluid streams move in opposite directions. The present problem can be formulated in terms of the boundary layer equations for two different heat transfer systems. These governing differential equations need to be considered separately and they are:

*Hot fluid*

$$\frac{\partial u_h}{\partial x} + \frac{\partial v_h}{\partial y} = 0 \qquad (1)$$

$$u_h \frac{\partial u_h}{\partial x} + v_h \frac{\partial u_h}{\partial y} = v_h \frac{\partial^2 u_h}{\partial y^2} - g\beta(T_h - t_h) \qquad (2)$$

$$u_h \frac{\partial T_h}{\partial x} + v_h \frac{\partial T_h}{\partial y} = \alpha_h \frac{\partial^2 T_h}{\partial y^2} \qquad (3)$$

where $u_h$ and $v_h$ denote the velocity components of the hot fluid in the $x$ and $y$ directions, respectively, $T_h$ is the temperature of the hot fluid, $g$ is the gravitational acceleration, $\beta$ is the thermal expansion coefficient of the hot fluid, and $v_h$ and $\alpha_h$ are the kinematic viscosity and thermal diffusivity of the heat fluid, respectively.

The boundary conditions for the free convection system are

$$\begin{aligned} u_h = v_h = 0, & \quad T_h = T_{wh}(x) \quad \text{on} \quad y = 0 \\ u_h \to 0, & \quad T_h \to t_h \quad \text{as} \quad y \to \infty \end{aligned} \qquad (4)$$

where $T_{wh}(x)$ denotes the wall temperature facing the hot side of the plate.

*Cold fluid*

$$\frac{\partial u_c}{\partial x_c} + \frac{\partial v_c}{\partial y_c} = 0 \qquad (5)$$

$$u_c \frac{\partial u_c}{\partial x_c} + v_c \frac{\partial u_c}{\partial y_c} = v_c \frac{\partial^2 u_c}{\partial y_c^2} \qquad (6)$$

$$u_c \frac{\partial T_c}{\partial x_c} + v_c \frac{\partial T_c}{\partial y_c} = \alpha_c \frac{\partial^2 T_c}{\partial y_c^2} \qquad (7)$$

where $x_c$ and $y_c$ are the Cartesian coordinates on the forced convection side, $u_c$ and $v_c$ denote the velocity components of the cold fluid in the $x_c$ and $y_c$ directions, and $T_c$, $v_c$ and $\alpha_c$ are the temperature, kinematic viscosity and thermal diffusivity of the cold fluid, respectively.

The boundary conditions for the forced convection system are

$$\begin{aligned} u_c = v_c = 0, & \quad T_c = T_{wc}(x_c) \quad \text{on} \quad y_c = 0 \\ u_c \to U_\infty, & \quad T_c \to t_c \quad \text{as} \quad y_c \to \infty \end{aligned} \qquad (8)$$

where $x_c = L - x$ and $T_{wc}(x_c)$ denotes the wall temperature facing the forced convection side.

Heat conduction along the plate is neglected in comparison with transverse heat conduction. The heat flux entering the left face of the plate is equal to that leaving the right face at any given vertical position $x$, i.e.

$$k_s \frac{T_{wh} - T_{wc}}{b} = -k_c \left.\frac{\partial T_c}{\partial y_c}\right|_{y_c=0} = k_h \left.\frac{\partial T_h}{\partial y}\right|_{y=0} = q_{xc} \qquad (9)$$

where $k_s$, $k_h$ and $k_c$ denote the thermal conductivities of the solid plate, hot fluid and cold fluid, respectively, and $q_{xc}$ is the local heat flux through the plate. A correlation between $T_{wh}$ and $T_{wc}$ can be obtained from equation (9) as

$$T_{wc} = T_{wh} - \frac{bk_h}{k_s} \left.\frac{\partial T_h}{\partial y}\right|_{y=0}.$$

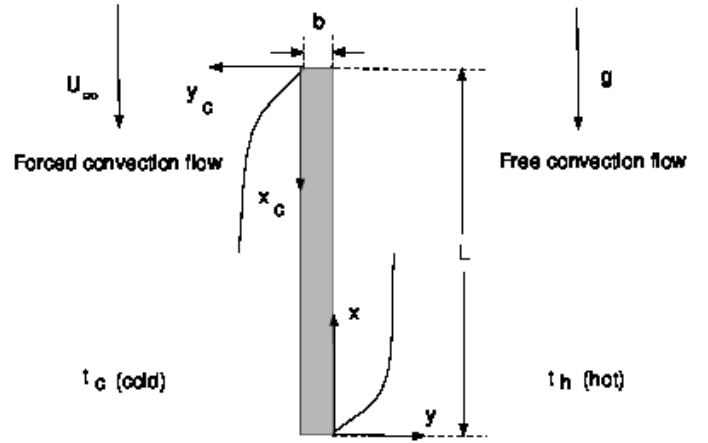

**Figure 1** Schematic diagram of physical model

## SOLUTION

To solve equations (1)-(3) and (5)-(7), the following dimensionless variables are introduced

$$\xi = \frac{x}{L}, \qquad \eta = \frac{C_h y}{L \xi^{\frac{1}{4}}}$$

$$\psi_h = 4\nu_h C_h \xi^{\frac{3}{4}} f_h(\xi,\eta), \qquad \theta_h(\xi,\eta) = \frac{T_h - \frac{t_h+t_c}{2}}{\Delta t}$$

$$\xi_c = \frac{x_c}{L} = 1 - \xi, \qquad \eta_c = \frac{y_c R_e^{\frac{1}{2}}}{L \xi_c^{\frac{1}{2}}}$$

$$\psi_c = (U_\infty \nu_c L \xi_c)^{\frac{1}{2}} f_c(\xi_c,\eta_c), \qquad \theta_c(\xi_c,\eta_c) = \frac{T_c - \frac{t_h+t_c}{2}}{\Delta t}$$

(10)

where $C_h = \left(\dfrac{g\beta\Delta t\, L^3}{4\nu_h^2}\right)^{\frac{1}{4}}$, $\Delta t = t_h - t_c$, $R_e = \dfrac{U_\infty L}{\nu_c}$ is the Reynolds number for the forced convection flow, and $\psi_h$ and $\psi_c$ are the stream functions of the hot and cold fluids, respectively, which are defined as

$$u_h = \frac{\partial \psi_h}{\partial y}, \qquad v_h = -\frac{\partial \psi_h}{\partial x}$$
$$u_c = \frac{\partial \psi_c}{\partial y_c}, \qquad v_c = -\frac{\partial \psi_c}{\partial x_c}.$$

(11)

Due to the definition of (10) and (11), equations (1)-(3) and (5)-(7) can be transformed into the following form

$$\frac{\partial^3 f_h}{\partial \eta^3} + 3 f_h \frac{\partial^2 f_h}{\partial \eta^2} - 2\left(\frac{\partial f_h}{\partial \eta}\right)^2 - \theta_h + \frac{1}{2}$$
$$= 4\xi \left(\frac{\partial f_h}{\partial \eta}\frac{\partial^2 f_h}{\partial \eta \partial \xi} - \frac{\partial^2 f_h}{\partial \eta^2}\frac{\partial f_h}{\partial \xi}\right)$$

(12)

$$\frac{1}{P_{rh}}\frac{\partial^2 \theta_h}{\partial \eta^2} + 3 f_h \frac{\partial \theta_h}{\partial \eta} = 4\xi\left(\frac{\partial f_h}{\partial \eta}\frac{\partial \theta_h}{\partial \xi} - \frac{\partial \theta_h}{\partial \eta}\frac{\partial f_h}{\partial \xi}\right)$$

(13)

$$\frac{\partial^3 f_c}{\partial \eta_c^3} + \frac{1}{2} f_c \frac{\partial^2 f_c}{\partial \eta_c^2} = \xi_c\left(\frac{\partial f_c}{\partial \eta_c}\frac{\partial^2 f_c}{\partial \eta_c \partial \xi_c} - \frac{\partial^2 f_c}{\partial \eta_c^2}\frac{\partial f_c}{\partial \xi_c}\right)$$

(14)

$$\frac{1}{P_{rc}}\frac{\partial^2 \theta_c}{\partial \eta_c^2} + \frac{1}{2} f_c \frac{\partial \theta_c}{\partial \eta_c} = \xi_c\left(\frac{\partial f_c}{\partial \eta_c}\frac{\partial \theta_c}{\partial \xi_c} - \frac{\partial \theta_c}{\partial \eta_c}\frac{\partial f_c}{\partial \xi_c}\right)$$

(15)

where $P_{rh} = \dfrac{\nu_h}{\alpha_h}$ and $P_{rc} = \dfrac{\nu_c}{\alpha_c}$ are the Prandtl numbers of hot and cold fluids, respectively.

The boundary conditions (4) and (8) become

$$f_h = \frac{\partial f_h}{\partial \eta} = 0, \qquad \theta_h = \theta_{wh}(\xi) \qquad \text{at } \eta = 0$$

$$\frac{\partial f_h}{\partial \eta} \to 0, \qquad \theta_h \to \frac{1}{2} \qquad \text{as } \eta \to \infty$$

$$f_c = \frac{\partial f_c}{\partial \eta_c} = 0, \qquad \theta_c = \theta_{wh}(\xi) - \frac{R_t R_t^*}{\xi^{\frac{1}{4}}}\frac{\partial \theta_h}{\partial \eta}\bigg|_{\eta=0} \qquad \text{at } \eta_c = 0$$

$$\frac{\partial f_c}{\partial \eta_c} \to 1, \qquad \theta_c \to -\frac{1}{2} \qquad \text{as } \eta_c \to \infty$$

where $R_t = \dfrac{bk_c R_e^{\frac{1}{2}}}{k_s L}$ denotes the thermal resistance ratio of the forced convection flow to the wall, $R_t^* = \dfrac{k_h C_h}{k_c R_e^{\frac{1}{2}}}$ can be regarded as the thermal resistance of the hot fluid to the cold fluid, $\theta_{wh} = \dfrac{T_{wh} - \frac{t_h+t_c}{2}}{\Delta t}$ and $\theta_{wc} = \dfrac{T_{wc} - \frac{t_h+t_c}{2}}{\Delta t}$. Substituting variables (10) into equation (9), it gets

$$R_t^*\left(\frac{\xi_c^{\frac{1}{2}}}{\xi^{\frac{1}{4}}}\right)\frac{\partial \theta_h}{\partial \eta}\bigg|_{\eta=0} + \frac{\partial \theta_c}{\partial \eta_c}\bigg|_{\eta_c=0} = 0 \quad \text{at any given position } x_c. \quad (16)$$

Based on equation (9), the local heat transfer coefficient $h_{xc}$ for the forced convection system can be expressed as

$$h_{xc} = \frac{-k_c \frac{\partial T_c}{\partial y_c}\bigg|_{y_c=0}}{T_c(x_c,0) - t_c} = \frac{q_{xc}}{T_c(x_c,0) - t_c}.$$

The local Nusselt number for the forced convection system can be expressed as

$$N_{uxc} = \frac{x_c h_{xc}}{k_c} = -\frac{R_e^{\frac{1}{2}} \xi_c^{\frac{1}{2}} \frac{\partial \theta_c}{\partial \eta_c}\bigg|_{\eta_c=0}}{\theta_{wc}(\xi_c) + \frac{1}{2}}.$$

The total heat flux $Q$ through the surface facing the cold fluid is obtained by integrating the local heat flux over the entire height of the plate and can be expressed as

$$Q = k_c \int_0^L \left(-\frac{\partial T_c}{\partial y_c}\bigg|_{y_c=0}\right) dx_c. \quad (17)$$

Substituting (10) into (17) yields the average Nusselt number for the forced convection system as

$$N_{uc} = \frac{Q}{k_c \Delta t} = R_e^{\frac{1}{2}} \int_0^1 \left(\frac{-\frac{\partial \theta_c}{\partial \eta_c}\bigg|_{\eta_c=0}}{\xi_c^{\frac{1}{2}}}\right) d\xi_c = A R_e^{\frac{1}{2}}$$

where

$$A = \int_0^1 \left(\frac{-\frac{\partial \theta_c}{\partial \eta_c}\bigg|_{\eta_c=0}}{\xi_c^{\frac{1}{2}}}\right) d\xi_c.$$

To solve equation (12) to (15), a comprehensive and non-iterative numerical scheme is proposed, which is in contrast with the iterative process purposed in [15,16] using a guessing strategy. Based on equation (16), the boundary conditions can be rewritten as

$$f_h = \frac{\partial f_h}{\partial \eta} = 0, \qquad \frac{\partial \theta_h}{\partial \eta} = \xi^{\frac{1}{4}} \lambda^*(\xi) \qquad \text{at} \quad \eta = 0$$

$$\frac{\partial f_h}{\partial \eta} \to 0, \qquad \theta_h \to \frac{1}{2} \qquad \text{as} \quad \eta \to \infty$$

$$f_c = \frac{\partial f_c}{\partial \eta_c} = 0, \qquad \frac{\partial \theta_c}{\partial \eta_c} = -R_t^* \xi_c^{\frac{1}{2}} \lambda^*(\xi) \qquad \text{at} \quad \eta_c = 0 \qquad (18)$$

$$\frac{\partial f_c}{\partial \eta_c} \to 1, \qquad \theta_c \to -\frac{1}{2} \qquad \text{as} \quad \eta_c \to \infty$$

and

$$\lambda = 0 \qquad \text{at} \quad \eta = 0, \; \eta_c = 0 \qquad (19)$$

where the dummy variable $\lambda$ is defined as

$$\lambda(\xi, \eta) = R_t R_t^* \lambda^*(\xi) - \theta_h + \theta_c \; . \qquad (20)$$

The systems (12)-(15) and (20), together with the boundary conditions (18) and (19), are then solved using the singular perturbation method and the difficulties associated with the guessed interfacial conditions have been obviated. Since this procedure is described in [13,14], it is not repeated here. Note that the points $\xi_c = 0$ and $1$ are singular which make the problem more difficult.

## RESULTS AND DISCUSSION

In this section the effects of the Prandtl numbers $P_{rh}$ and $P_{rc}$, and resistance parameters $R_t$ and $R_t^*$ on the interface temperatures, heat transfer rates and Nusselt numbers are discussed. Figures 2 and 3 show the variation of $\theta_{wc}(\xi_c)$ with $\xi_c$ for some values of $R_t$ and $R_t^*$ when the two working fluids have $P_{rh} = P_{rc} = 1$. The results show that for the increasing values of $R_t$ the temperature of the cold side of the wall decreases. It happens because when $R_t$ is increased the wall becomes more effective insulation between the two forced and free convection flows. In contrast, $\theta_{wc}(\xi_c)$ increases as the free convection becomes dominant, i.e. when the parameter $R_t^*$ increases. It turns out that an increase of $R_t$ leads to a reduction of the heat transfer rates at the cold side of the wall and to an increase of these rates with increasing $R_t^*$, as can be seen from Figures 4 and 5. It is evident from Figures 6 and 7 that the effects of $R_t$ and $R_t^*$ on the local Nusselt number are insignificant. To this end it should be mentioned that the same trends persist for the heat transfer characteristics of the free convection system [17-23].

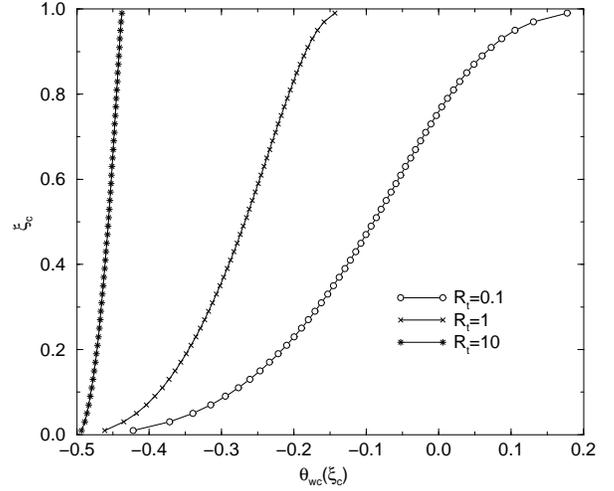

**Figure 2** Effect of $R_t$ on $\theta_c\big|_{\eta_c=0}$ for $R_t^* = P_{rh} = P_{rc} = 1$

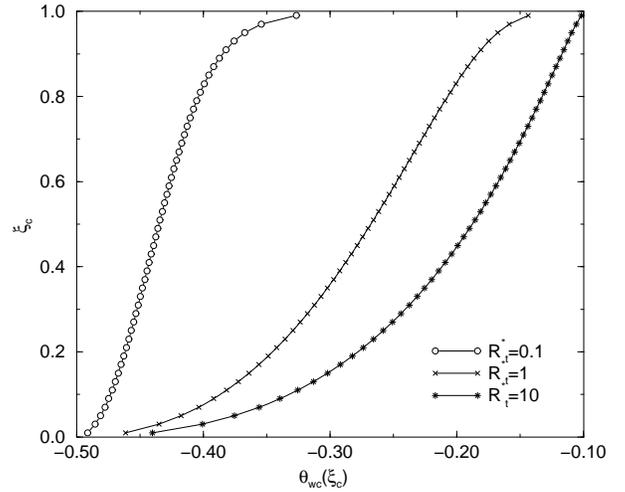

**Figure 3** Effect of $R_t^*$ on $\theta_c\big|_{\eta_c=0}$ for $R_t = P_{rh} = P_{rc} = 1$

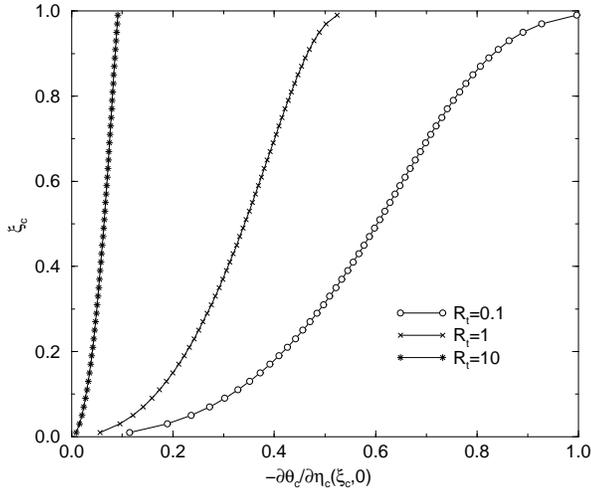

**Figure 4** Effect of $R_t$ on $-\left.\dfrac{\partial \theta_c}{\partial \eta_c}\right|_{\eta_c=0}$ for $R_t^* = P_{rh} = P_{rc} = 1$

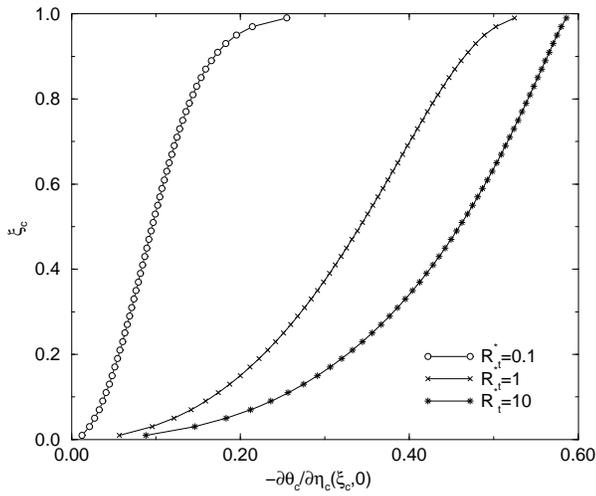

**Figure 5** Effect of $R_t^*$ on $-\left.\dfrac{\partial \theta_c}{\partial \eta_c}\right|_{\eta_c=0}$ for $R_t = P_{rh} = P_{rc} = 1$

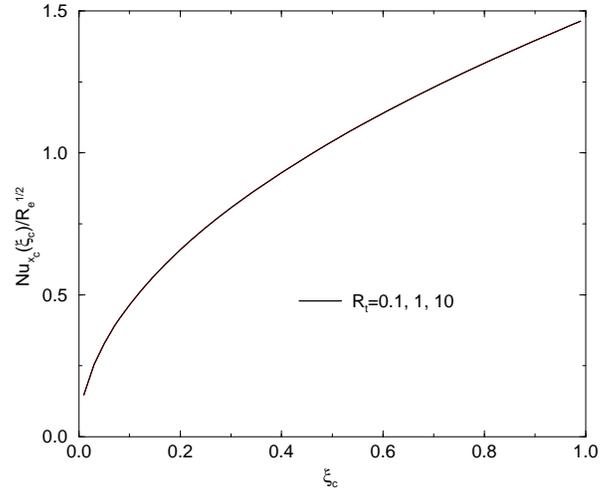

**Figure 6** Effect of $R_t$ on $\left.\dfrac{N_{uc}}{R_e^{\frac{1}{2}}}\right|_{\eta_c=0}$ for $R_t^* = P_{rh} = P_{rc} = 1$

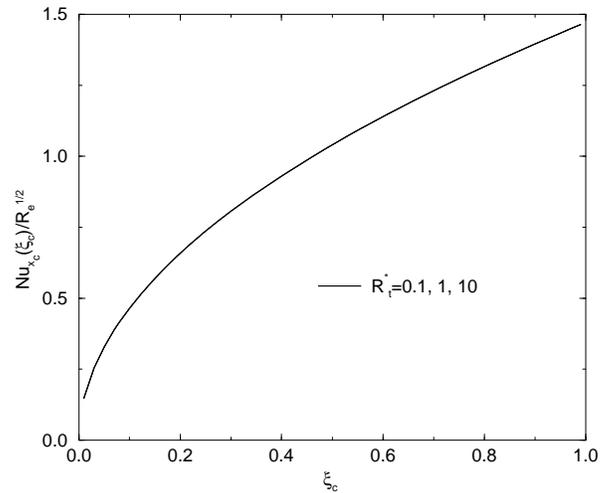

**Figure 7** Effect of $R_t^*$ on $\left.\dfrac{N_{uc}}{R_e^{\frac{1}{2}}}\right|_{\eta_c=0}$ for $R_t = P_{rh} = P_{rc} = 1$

## CONCLUSION

A conjugate problem of heat transfer between a laminar forced convection flow and a laminar free convection separated by a vertical finite wall is studied theoretically. The axial thermal conduction in the wall is neglected. The governing boundary layer equations subject to conjugate boundary conditions are solved numerically using a very efficient method which differs from the one used by other authors. As a result the temperature distributions and heat transfer rates at both sides of the wall have been determined. The results show that the resistance parameters influence substantially the interactive heat transfer characteristics. A particular attention is given to the situation in which the fluids at both sides of the wall are the same. It is worth mentioning that the results reported in this paper are in general in agreement with those from the open literature. However, the accuracy of these results can be further evidenced through experiments.